# TOWARDS SELF-OPTIMIZATION OF PUBLISH ∕ SUBSCRIBE IOT SYSTEMS USING CONTINUOUS PERFORMANCE MONITORING

Djahafi Mohammed and Salmi Nabila

MOVEP Laboratory, USTHB University, Bab-Ezzouar, 16111, Alger, Algeria

*ABSTRACT*

*Today, more and more embedded devices are being connected through a network, generally Internet, offering users different services. This concept refers to Internet of Things (IoT), bringing information and control capabilities in many fields like medicine, smart homes, home security, etc. Main drawbacks of IoTenvironment are its dependency on Internet connectivity and need continuous devices power. These dependencies may affect system performances, namely request processing response times. In this context,we propose in this paper a continuous performance monitoring methodology, applied on IoT systems based on Publish/subscribe communication model. Our approach assesses performances using Stochastic Petri net modeling, and self-optimizes whenever poor performances are detected. Our approach relies on a Stochastic Petri nets modelling and analysis to assess performances. We target improving performances, in particular response times, by online modification of influencing factors.*

## 1. INTRODUCTION

Twenty years ago, Internet of things was introduced as a network of "things" endowed with sensor technology and ability to communicate. Today, things include physical devices like smartphones, home security systems, smart home appliances, printers and webcams, medicine peripherals, etc. The main objective of IoT is to create an interconnected system of devices that can automate processes and improve efficiency in various industries and daily life. For instance,in medicine, medical professionals' patients monitoring may be done thanks to smart devices connecting patients inside and outside a hospital, allowing practitioners to adjust treatments timely at the required time, hence saving and improving patient health.

To achieve that, several communication protocols have been developed, such as MQTT (Message Queuing Telemetry Transport) [1]. This latter is a lightweight protocol designed specifically for IoT devices with limited processing power and bandwidth. Small amounts of data are exchanged through MQTT between connected devices, enabling consequently efficient and reliable communication, even in low-bandwidth or unreliable network conditions. In fact, MQTT relies on a publish/subscribe pattern [2], which allows multiple clients to receive messages simultaneously.

A challenging issue in IoT environments is to maintain a good performance level, in particular, improved response times. A response time refers to the time period it takes for a message to be sent from one device to another. However, performances may be affected by several factors, including connection speed, distance between devices, processing devices power, etc. For example, in a smart home system, a poor response time may lead to situations where lights do not turn on/off when needed, or the temperature control does not function correctly. To ensure optimal functioning, it is essential to be able to analyze performances and act whenever needed. For this aim, we propose to endow IoT systems with continuous performance monitoring ability, where key parameters affecting performances can be adjusted adequately in real-time. These parameters include network latency, processing power, memory, and software optimization. By monitoring and adjusting these parameters, IoT systems can achieve faster response times and improve overall performances.

In this context, we present in this paper a performance monitoring methodology, based on Stochastic Petri nets (SPN) [19] performance analysis and applied on a Publish/subscribe IoT system. The approach consists of building an SPN model of a current system configuration and predicting performance to detect degraded situations. We try to bring out parameters influencing response times.

The proposed performance monitoring methodology relies on SPN as a key modeling technique, with a primary focus on capturing the probabilistic behavior exhibited by IoT systems. In fact, as IoT environments are characterized by uncertainties and variabilities in data, communication, and user interactions, it becomes crucial to employ modeling approaches to effectively represent and analyze these probabilistic elements. SPNs offer a well-suited formalism for capturing and representing the stochastic nature of IoT systems, making them an essential tool in understanding and optimizing performances of such systems.

This paper is organized as follows: Section 2 discusses related work. Section 3 presents the publish/subscribe model. Section 4, describes our performance monitoring methodology. Experimental results are then presented in Section 5. Finally, Section 6 concludes the paper.

## 2. RELATED WORK

Several studies have been conducted on analyzing and optimizing the performance of Internet of Things (IoT) systems.

Duran et al. [9] present an automated approach to compose IoT applications from available devices based on user goals. This study used labeled transition systems to formally model device behaviors, which enables synthesis of customized IoT apps matching user needs with minimal human effort.

Garofalaki et al. [24] developed an ADOxx-based modeling tool called SAPnet that enables security evaluation of Internet of Things (IoT) systems modeled as stochastic Petri nets by associating vulnerabilities to model states and quantifying security metrics for assessment during the design phase. Elleuch et al. [4] present a formal analysis approach using higher-order logic theorem proving to check coverage properties and performance of wireless sensor networks. It employed randomize node scheduling, and apply it to an IoT-based environmental monitoring framework.

Hwang et al. [5] suggested a federated learning local control study for IoT home services, aiming at minimizing end-to-end response times. Besides, Herrera et al. [6] proposed DADO framework, which provides scalable deployment plans that trade-off execution time and reduce response time. Another measurement study, conducted by Lee et al. [7], examines common IoT home devices response times, such as smart lights and smart plug. This study includes a performance comparison with and without cloud concept.

Besides, an interesting survey on Publish/Subscribe systems by Lazidis et al. [13] highlights design features and technologies using real-case scenarios for performances evaluation. Furthermore, Ferraz Junior et al. [14] discuss the need for IoT messages to achieve both energy efficiency and secure delivery, and propose a standardization for the topic and payload for publish-subscribe systems. They evaluate performances of such systems, including AMQP, DDS, and MQTT, using their proposed energy-efficient message structure. Moreover, Sreeraj et al. [17] proposed a prediction framework for Publish/subscribe protocols within restricted wireless access network. Docker containers performances were also studied by MORABITO [18] by using bench-mark tools, when running on a single board computer device such as Raspberry Pi 2. Also, main network performance issues in an IoT system are presented by Sowmya et al. [8].

With regards to MQTT brokers in IoT, multiple studies evaluated their performances. Bertrand-Martinez et al. [10] proposed a methodology that involves classification and evaluation of IoT Brokers. Similarly, de Oliveira et al. [11] analyzed MQTT brokers performance regarding latency during data packets transport. Another study assessing performances of several MQTT broker implementations using stress testing was proposed by Mishra et al. [12]. Also, Kashyap and Sharma [15] studied performances of Stimulus Systems they designed using an MQTT broker. Furthermore, Duttagupta et al. [16] predicted performances of specific APIs offered by an IoT platform using queuing network modelling.

For our part, we target to introduce self-optimization to contribute with aforementioned proposals in improving IOT systems performances. In this objective, we present, in this paper, a continuous performance monitoring approach for self-optimizing publish/subscribe IoT systems, based on Stochastic Petri net modelling. The motivation behind using SPN modeling is to address inherent complexity and dynamic nature of IoT systems, as well as capturing concurrency, synchronization, and stochastic aspects. Notice that traditional performance monitoring methods often struggle to capture intricate interactions and stochastic behavior of IoT systems, hindering the optimization process. So, we can efficiently represent and analyze publish/subscribe IoT systems behavior, considering inherent uncertainties and variability.

## 3. PUBLISH/SUBSCRIBE MODEL

The publish/subscribe model [2] defines a loosely coupled communicating pattern, to reach easy improved communication and increase reliability and scalability. This paradigm is commonly used in the context of IoT systems to facilitate communication between devices, sensors, and applications. Two entities are defined: publishers sending messages to a central message broker, and subscribers which declare their interest in receiving messages of a specific "topic". The broker distributes received messages to appropriate subscribers according to their interests. It allows devices to communicate with a wide range of applications and services without the need for direct integration, which simplifies development and maintenance. Additionally, the publish-subscribe model can help in reducing network congestion and improving overall system performance by allowing applications to only receive the needed data. A lightweight protocol that implements the publish/subscribe model is MQTT [1] (Message Queuing Telemetry Transport), which is designed to be efficient, reliable, and suitable for constrained devices with limited processing power, memory, and bandwidth.

## 4. PERFORMANCE MONITORING METHODOLOGY

### 4.1. Key idea

Our purpose is to propose a self-optimization approach enhancing IoT systems performances. The approach consists of detecting poor performance situations and trying to remedy such situations by online modification of causal factors. In this paper, we target response time improvement for Publish/subscribe IoT systems where devices publish messages and others are notified with published messages. Our methodology is based on SPN modelling and analysis.

### 4.2. SPN Model

A stochastic Petri net is a formal model used to describe the behavior of concurrent systems, incorporating both discrete and probabilistic aspects.

Definition: A Stochastic Petri Net [19,20,21] is a couple (N, R) defined with:

- N= (P; T; Pre; Post; Inh; Pri) is a Petri Net [22] where: - P is a set of places.
— T a set of timed transitions with a stochastic firing delay.
— Pre, Post and Inh are respectively precondition, postcondition and inhibition functions relating transitions to places.
— Pri is the transition priority function.
- R = {r1, r2, ...rm} is a set of firing rates, associated to a timed transition, m is the cardinality of T.

A state of a SPN (said marking), is a vector giving for each place the number of contained marks.

The motivation for using SPN modelling in our approach is its ability to represent the complex dynamics of IoT systems, considering concurrency, conflicts and probabilistic behavior. Thanks to its isomorphism with queuing network, we are able to compute system performance metrics, such as response time and resource utilization, and identify performance bottlenecks. So, real-time monitoring based on SPN based performance prediction enables detection of degraded situations, allowing for timely actions to improve system performance.

### 4.3. IoT Performance Monitoring Process

Our methodology follows three steps (see Figure 1):

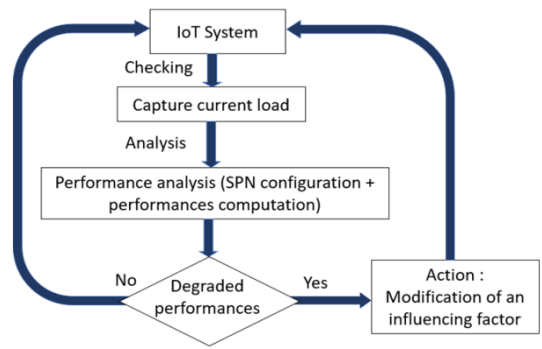

Figure 1. Performance monitoring methodology.

**Step 1.** Capture, within the IoT system, current load to get main configuration parameters to be used in our model. An of such parameter is requests number.

**Step 2.** Build an SPN model configured with captured load parameters, and compute performance metrics like response times of submitted requests.

**Step 3.** In case of performance degradation, apply predefined actions to get rid of poor performances. These actions consist of changing one of main factors impacting response times.
The process depicted in Figure 1 is iterative, meaning that it loops back to the monitoring phase after optimization actions have been applied. This allows for continuous performance monitoring and real-time adjustments to ensure optimal functioning of the IoT system.

### 4.4. Performance Influencing Factors

There are several factors that may affect IOT systems performances, and in particular response times. We give below main factors.

#### 4.4.1. Broker Memory

It refers to the amount of memory allocated to the broker process for storing and processing messages. In fact, the required memory amount depends on the subscribers and publishers number, the number of submitted requests being processed and the messages size.

#### 4.4.2. Network Buffering

It defines memory buffers used to receive or send data from or to IoT devices, i.e., publishers and subscribers. Memory buffers are used as well on broker servers as on network devices through which data is routed. This parameter may have a significant impact on response times. For instance, ActiveMQ [3] broker relies on TCP for communication. To optimize response times, we can change TCP buffers size or settings.

#### 4.4.3. QoS Level

Generally, broker software uses a concept of quality of service to configure messaging transport. Several QoS values are usually defined to offer users an appropriate QoS. However, the more the QoS level is incremented, the more response times get increasing, as more processing time is required.

### 4.5. SPN Modelling

In this paper, we followed an approach by example to define our modeling approach. For this aim, we study an MQTT Publish/Subscribe based IOT platform. This platform consists of three key elements: the broker, publisher, and subscriber. To effectively model the publish/subscribe platform, we adopt a modular approach: each component, including the client connection, subscription, publication and notification, is modeled separately as sub-models. This allows us to focus on component specificities and behavior. Interactions between components are then represented by connecting sub models together.

We also need to highlight, in our model, influencing factors impact on the IoT platform. So, we model dependencies related to these factors, as they play crucial roles in IoT system performances and reliability. By incorporating these factors into the SPN model, we can assess and predict their impact on performance metrics. This enables us to understand trade-offs and make adequate decisions regarding the configuration and optimization of the IoT platform.

Note that, in our SPN modeling, we abstract internal implementation details regarding publication and subscription management to a given topic. Our goal is to capture the overall system behavior in terms of performance, rather than implementation matters. As such, our model focuses on incoming events such as the arrival of a publish or subscribe request, and outgoing events like sending notifications to subscribers.

### 4.5.1. Client Connection/Disconnection sub-model

Figure 2 represents clients behavior (for publishers or subscribers). This sub-model focuses on the client's connection and disconnection processes. It models a *"publisher"* or *"subscriber"* through a place containing client instances as its marking. In order for a client to publish an event or subscribe to a topic, it must first establish a connection with the broker server. This connection process is captured by the *"connect"* transition within the sub-model. Once a client initiates a connection request, the broker, represented by the *"broker"* place, checks if the request can be accepted according to its available resources. The acceptance of the connection request is modeled by the *"Accept"* transition, which takes into account the broker capacity (*"BrokerCapacity"* place), modelling the maximum number of clients that can be connected simultaneously to the broker. If the connection request is accepted, the client token moves from the *"publisher"* or *"subscriber"* place to a new place representing a connected client (publisher or subscriber).

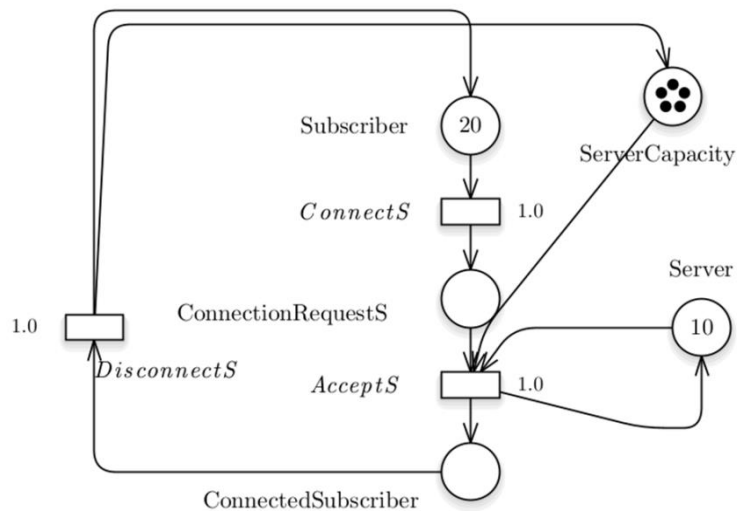

Figure 2. Connection/disconnection

### 4.5.2. Subscription/Unsubscription sub-model

This sub-model captures specifically subscription and unsubscription processes for connected subscribers, (see Figure 3). At the core of this sub-model is a dedicated place named "*Topics*" that represent possible topics to which subscribers can subscribe. The subscription process is initiated by the "*subscribe*" transition. This transition establishes the association between the subscriber and the desired topic, enabling them to receive related messages. Importantly, subscribers also have the flexibility to unsubscribe from topics (via the "*unsubscribe*" transition) when they no longer wish to receive associated messages.

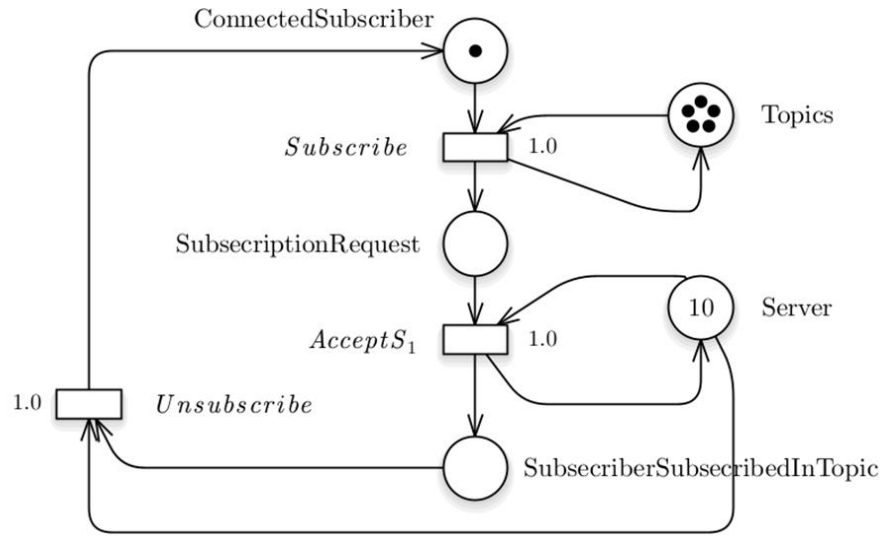

Figure 3. Suscription sub-model.

### 4.5.3. Publication sub-model

Sub-model of Figure 4 represents the publication process for connected publishers. Mainly, we have the "*publish*" transition, which is triggered when a connected publisher has an event or message to publish. The events or messages to be published are represented by tokens in the "*EventToPublish*" place. The publication request is then processed by the broker, which check its validity and capacity to handle the publication. This acceptance action, denoted by the "*AcceptPub*" transition, indicates that the broker has acknowledged the publication request and has queued it in its requests queue. To accurately model the broker's memory constraints, we introduce the "*BrokerMemory*" place, representing the available memory resources within the broker. This ensures that the publication process takes into consideration the maximum capacity for storing published data    . Additionally, we incorporate the "*NetworkReceiveBuffer*" place, which represents the buffer space allocated for incoming network data. This buffer is used during reception/sending of published data within the network infrastructure. Furthermore, to consider the Quality of Service (QoS) level for publications, we include the "*PubQoSProcessing*" transition. This transition reflects the processing performed by the broker to ensure adherence to the specified QoS level for each publication, involving actions such as acknowledgment handling or message duplication for reliable delivery. By integrating these features into our publication sub-model, we obtain a faithful representation of the publication process, highlighting resource utilization, network considerations and QoS compliance, enabling us to analyze impact of influencing factors on the IoT system.

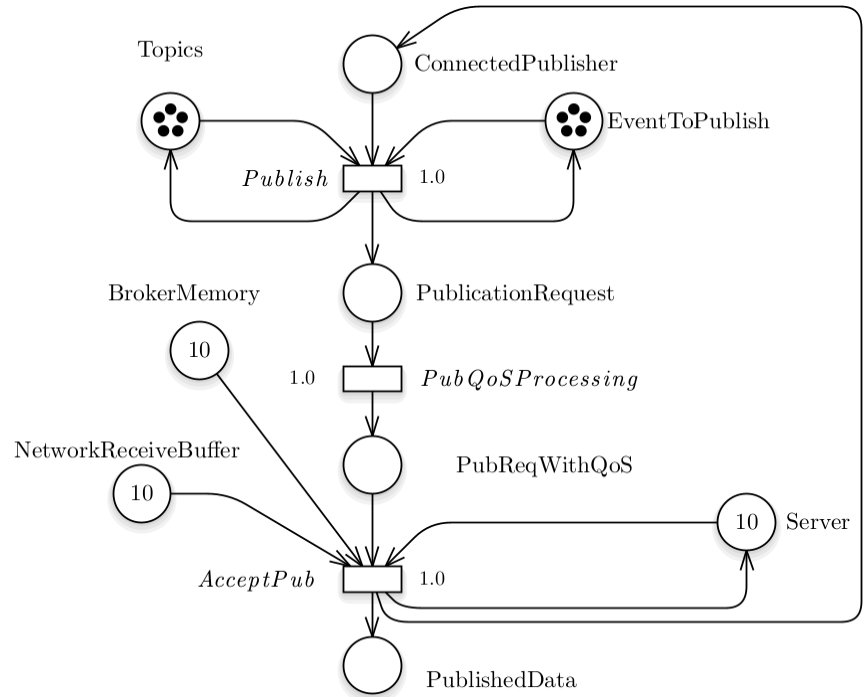

Figure 4. Publication sub-model.

### 4.5.4. Subscriber notification sub-model

Once a message is published for a specific topic, the broker is responsible for notifying subscribers who have subscribed to the same topic, (see Figure 5). This notification process (represented by "*notify*" transition), consists of delivering the published message to interested subscribers. Upon receiving the notification, the subscriber consumes the received event/message, which is represented by the "*consume*" transition. The consumption process indicates that the subscriber has successfully received and processed the event/message. To consider memory constraints associated with receiving published data, we have introduced the "*ReceivedEventCapacity*" place. This place represents the available subscriber capacity to store received events. It ensures that the system considers the maximum capacity for storing and handling incoming events, preventing thus memory overflow or resource congestion. Furthermore, we have introduced the "*SubQoSprocessing*" place to reflect quality of service processing for handling subscriber, QoS specificities before notification. We have further included the "*NetworkSendBuffer*" place that represents the buffer capacity within the system for outgoing network transmissions. When a message is being sent from the broker to the subscribed subscriber, it is temporarily stored in the network send buffer before being transmitted over the network.

### 4.6. Global Model

Once the sub-models were developed, we merge communicating transitions for interconnecting each sub model with the other to obtain the overall SPN model representing the entire IoT publish/subscribe platform. Each sub-model captures specific system behavior, and their composition enables modeling the full system (see Figure 6).

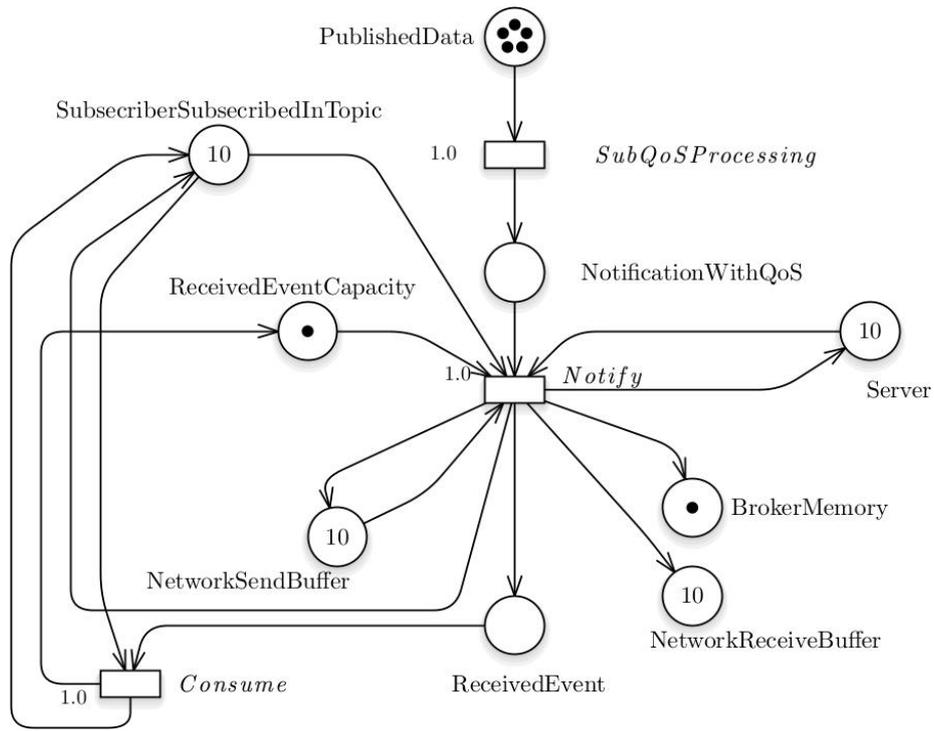

Figure 5. Notification sub-model.

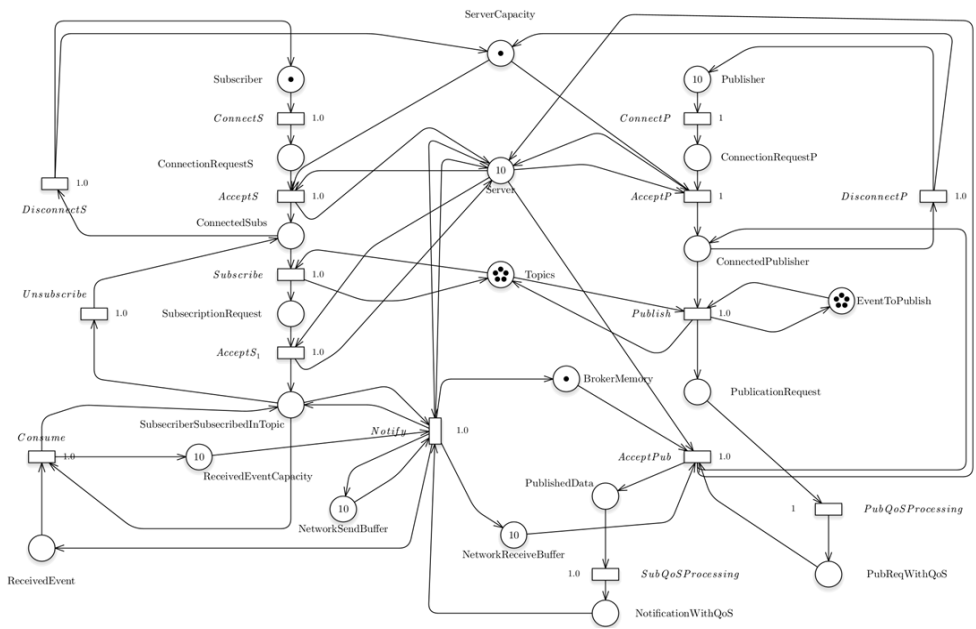

Figure 6. SPN model of the entire publish/subscribe platform.

## 5. EXPERIMENTAL RESULTS

To assess MQTT IoT platform performances, we conducted a series of experiments using our global SPN model. In our monitoring process, we propose to perform an action to optimize performances of the platform when a performance degradation is detected. In our case, we defined possible optimization actions as those strengthening resources related to influencing factors presented in section 5.2. Hence, we present in this section experiments bringing out impact of factors mentioned above. For each experiment, we varied one influencing factor while keeping the others constant, allowing us to isolate the effects of each factor on the system's performance metrics. We used GreatSPN [23] for performance analysis of our global model: computation of steady state probabilities and performance metrics.

### 5.1. Network Buffering Impact

We investigated in this experiment the network buffering impact on system performances. In practice, this parameter is handled by varying the network send and receive buffer size, hence modifying in our model, the "*NetworkReceiveBuffer*" and "*NetworkSendBuffer*" places content. We assessed performances, namely publication acceptance and subscriber notification response times. These investigations may help in choosing the optimal buffer size required to get improved system performances, while balancing network resources (see Table 1).

Table 1. Network buffering impact on response time.

| Influencing factor | Network buffering | |
|---|---|---|
| Buffer size (in terms of requests number) | 1 | 10 |
| Accept Publication response Time | 8,50740309 | 6,09202230 |
| Notification response Time | 9,06444339 | 6,21357193 |

This table shows that increasing network buffering size has a positive impact on response times. This highlights the importance of considering network buffering as a critical factor for optimizing system performance.

### 5.2. Broker Memory Impact

We explore in this experiment the broker memory impact on system performances. We simulated different memory capacities by varying in our model the "*BrokerMemory*" place marking, and observed how it affects the system ability to handle concurrent connections. This analysis provided insights into the optimal memory allocation, to ensure smooth request processing under various workloads (see Table 2).

Table 2. Memory size impact on response time.

| Influencing factor | Broker Memory | |
|---|---|---|
| Memory size (in terms of requests number) | 1 | 10 |
| Accept Publication response Time | 8,50740309 | 6,09202230 |
| Notification response Time | 9,06444339 | 6,21357193 |

### 5.3. QoS Level impact

We conducted this experiment to show efficiency of QoS modification action, which can be applied to an IoT system whenever performances are degraded. Hence, we show how performances may be degraded when QoS processing rates increase.

For this purpose, we vary, in our SPN model, we vary the "*PubQoSProcessing*" transition rate (see Figure 7) and compute the broker response time needed for accepting and processing a publication (for publishers) and response time required for notifying a subscriber. For critical applications requiring high QoS, reducing QoS is generally not a viable option. We note that the response times are measured in units of time, such as microseconds (ms), representing the average time taken to process publish and notification requests based on the SPN model simulations. The connection rate is measured in number of connections per unit, indicating the rate at which client publications are established per units of time.

In Figure 7, we observe how the "*PubQoSProcessing*" transition rate affected publication response times, as well as subscribers' notification response times. By reducing QoS level in a broker, we improve processing times. The higher are processing rates, the more rapid publication and notification can be processed, resulting in reduced response times for both publishers and subscribers. This outcome indicates that the platform's performance, in terms of delivering notifications, is being improved with higher "*PubQoSProcessing*" transition rates. While decreasing the QoS level can improve response times as shown in our results, this action is only appropriate for some less critical applications. For systems requiring high QoS, other factors like network buffer size should be tuned instead.

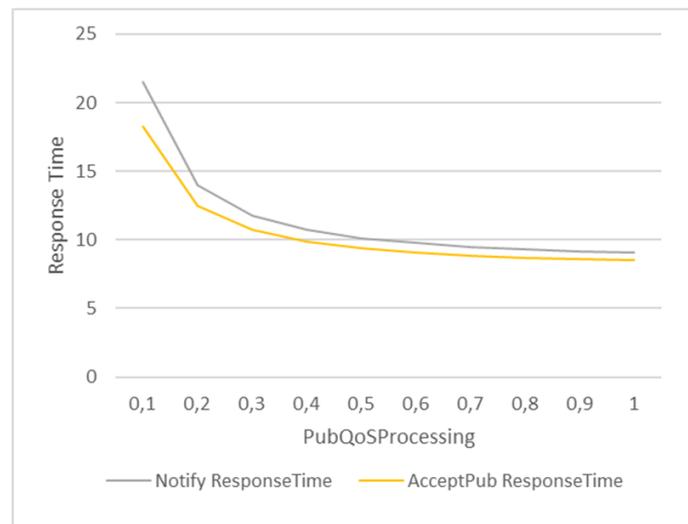

Figure 7. Response time compared to PubQoSProcessing rate.

## 5.4. Summary

Through our experimentation, we have shown how our monitoring process can monitoring process can detect poor performances, and adequately choose an improved solution by modifying one of the influencing factors and checking performances of the resulted system configuration. Consequently, the monitoring process will enable self-optimization to achieve desired performance objectives and meet specific IoT application requirements.

## 6. CONCLUSION

In this paper, we have presented a performance monitoring methodology to enable self-optimization of publish/subscribe IoT systems. Our approach leverages Stochastic Petri Net (SPN) modeling and analysis to predict performance metrics and identifies influencing factors on response time.

To present our method, we followed an approach by example, by studying performance analysis of an MQTT publish/subscribe IOT platform, as the publish/subscribe model offers efficient, reliable, and suitable communication protocol for constrained devices like IoT devices.

A key advantage of our methodology is the modular modeling approach based on sub-models. We developed sub-models representing the different components and behaviors of the MQTT publish/subscribe IoT platform, including client connection, subscription, publication, and notification. These sub-models were then interconnected to form the overall SPN model of the system. By composing reusable sub-models, we could accurately capture the dynamics and performances of the complete IoT platform.

We conducted experiments to demonstrate how our SPN model can be used to monitor factors like QoS level, network buffering, and broker memory utilization. The results showed that by detecting degraded performance and modifying influencing factors, our approach can significantly improve response times. This enables real-time optimization and self-adaptation of IoT systems to meet desired performance objectives.

We want to extend this work by providing a set of tools generating models of such IoT platforms and performance analysis for detecting performance degradation. An action will be triggered by a healing tool to configure adequately the studied IOT platform, so that to gain in performances. Another future work is to expand considered factors and self-optimize for different application requirements. The proposed approach can be leveraged to build self- adaptive IoT systems that can maintain high performance and efficiency.